\documentclass{elsart}

\usepackage{epsf}

\usepackage{amssymb}

\begin{document}
\begin{frontmatter}



\title{On the power spectrum of magnetization noise}


\author[IEN]{G. Durin\corauthref{cor1}},
\corauth[cor1]{Corresponding author. Phone: +39 011 3919751 Fax:
+39 011 3919782} \ead{durin@ien.it}
\author[Sap]{S. Zapperi}

\address[IEN]{Istituto Elettrotecnico Nazionale Galileo Ferraris and INFM, Corso M.
d'Azeglio 42, I-10125 Torino, Italy}
\address[Sap]{INFM sezione di Roma 1, Dipartimento di Fisica, Universit\`a "La
Sapienza", P.le A. Moro 2, 00185 Roma, Italy}

\begin{abstract}
Understanding the power spectrum of the magnetization noise is a
long standing problem. While earlier work considered superposition
of 'elementary' jumps, without reference to the underlying
physics, recent approaches relate the properties of the noise with
the critical dynamics of domain walls. In particular, a new
derivation of the power spectrum exponent has been proposed for
the random-field Ising model. We apply this approach to
experimental data, showing its validity and limitations.
\end{abstract}

\end{frontmatter}
\newpage
After almost a century since its discovery in 1909, the
magnetization noise produced by the intermittent motion of domain
walls, i.e. the Barkhausen noise, still represents an intriguing
scientific challenge from the theoretical point of view.
Considering the long-time and vast production of experimental and
theoretical papers of the past, it is quite surprising that only
recently an exhaustive comprehension of the noise properties has
been achieved. The introduction of methods of statistical
mechanics, in fact, made possible a reliable description of the
intrinsic complexity of magnetization processes. In particular,
the power law exhibited by the Barkhausen signal amplitude
together with the avalanche size and duration has been explained
in terms of an underling critical point. Its true nature is still
under debate, as two main different approaches have been proposed:
the zero temperature random-field Ising model (RFIM) \cite{SET-01}, where
criticality is set by the amount of disorder, and interface model,
where a domain wall moves through a disordered medium and
criticality is due to the depinning transition of the wall
\cite{CIZ-97,ZAP-98,DUR-00}. Exploiting the effects of long range
dipolar magnetostatic fields and of domain wall elastic tension on
the depinning transition, we have recently shown that two distinct
universality classes exist, with different critical exponents and
cutoff scaling dependence on the demagnetizing factor
due to sample geometry \cite{DUR-00}. These results have been
confirmed on two polycrystalline and amorphous sets of materials,
supporting a domain wall theory for the Barkhausen effect.

Despite these significant results, a proper description of the
shape of the power spectrum noise is still not available.
Significantly, about 80\% of the Barkhausen literature have been
devoted to this problem. Earlier approaches considering a
description of the power spectrum shape as a superposition of
elementary independent events (see for instance
\cite{MAZ-62b,ARQ-68}) appeared to be unrelated to any microscopic
mechanism, thus not clarifying the true origin of the
magnetization process. Even attempts to link the power spectral
exponent to the critical exponent of size distribution by a simple
scaling relation \cite{LIE-72,DAH-96,SPA-96} appears to be quite
unsatisfactory and not confirmed in general by experiments.
Experimental noise in fact show a quite complex pattern: the high
frequency part often follows a power law with exponents between
1.5 and 2, even if cases have been reported where the power law
extension is very limited and a more complex pattern results
\cite{DUR-96}. The low frequency part displays a marked peak,
those position strongly depends on the magnetization rate, and a
$f^\alpha$ dependence at lower frequencies, with $0.5 < \alpha <
1$. Such a complicated pattern does not actually reveal all the
complexity of the underlying dynamics of Barkhausen avalanches.
Considering high order moments of the signal, different frequency
bands appear to be very strong coupled, so that also high order
power spectra display power law dependences \cite{PET-98a,PET-98}.
In addition, time asymmetries of third-order voltage correlations
are found in amorphous samples, showing as high frequency events
precede on average the low frequency ones \cite{PET-98a}. All
these properties are not currently explained by any of the
existing models.

A derivation of the power spectrum exponent from a
scaling analysis, at least for some simple models, can help for a
better comprehension of the avalanche dynamics. A decisive step in
this direction has been performed recently by Kuntz and Sethna
\cite{KUN-00}, who derived the power spectrum exponent in the zero
temperature RFIM, and corrected the
earlier theoretical estimations \cite{LIE-72,DAH-96,SPA-96}. Most
part of their derivation can be applied to different models, and
also to the analysis of real data. In this paper, we apply some of
their results to various sets of experimental data, confirming the
general validity of this approach for simple cases, and showing
where it must be improved for a better description of the results.

Let us consider some of the relations of Ref.~\cite{KUN-00} that
we will use in our analysis. The key scaling relation connects the
average avalanche size $\langle s(T)\rangle$ with its duration $T$, that is
$\langle s(T)\rangle\sim T^{1/\sigma \nu z}$ where the exponents
$\sigma$, $\nu$ and $z$ are defined in Ref.~\cite{DAH-96}.
When the exponent of the avalanche
size distribution $\tau$ is less than 2, as usually in experiments
\cite{DUR-00}, the high frequency tail of power spectrum is
calculated to scale with exponent ${1/\sigma \nu z}$. This central
result is based on the existence of a couple of scaling relations
regarding the avalanche shape. The first one states that the
average avalanche shape should scale in a universal way, so that
\begin{equation}\label{Vt}
  V(T,t)=T^{1/\sigma \nu z-1}f_{shape}(t/T)
\end{equation}
where $V$ is the signal voltage, $t$ is the time and
$f_{shape}(t/T)$ is a universal scaling function, having the
approximated shape of an inverted parabola for the RFIM. The
second relation analyzes the fluctuations of avalanche sizes
considering the probability $P(V|s)$ of the occurrence of voltage
$V$ inside an avalanche of size $s$. This probability scales as:
\begin{equation}\label{PVS}
  P(V|s)=V^{-1} f_{voltage}(Vs^{\sigma \nu z-1})
\end{equation}
where $f_{voltage}$ is another universal scaling function. With
relations \ref{Vt} and \ref{PVS}, the power spectrum exponent is
obtained calculating the time-time correlation function in the
case of adiabatically increase of the applied field and of a {\em
complete separation of avalanches in time}, thus avoiding any
avalanche correlation. All these considerations will be helpful to
understand the experimental results.

\begin{figure}
\centerline {        \epsfxsize=7.5cm
                \epsfbox{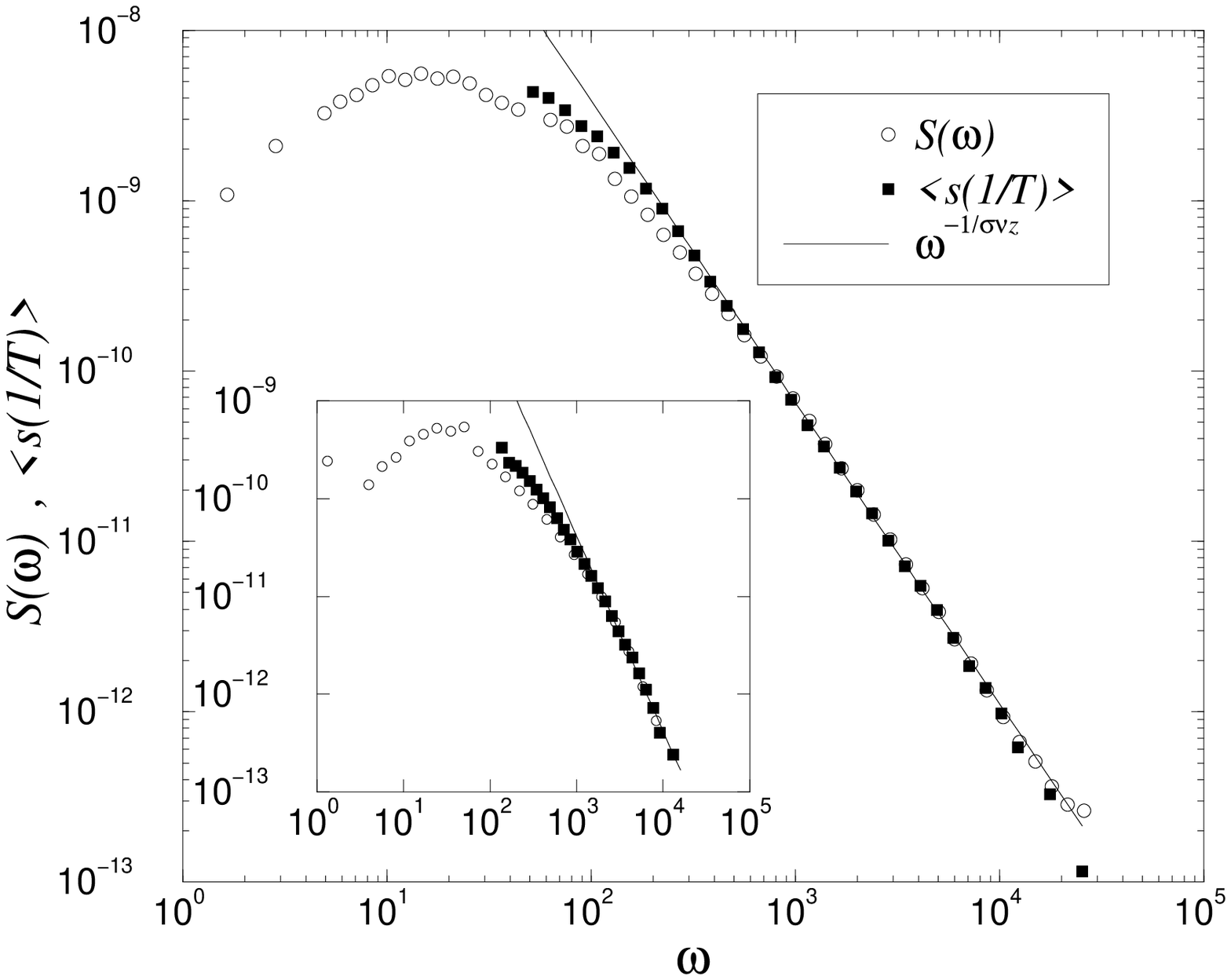}
        }
\caption{Comparison of the power spectrum $S(\omega)$ with the
average avalanche size $s(1/T)$ as a function of inverse of
avalanche duration $T$ for an $Fe_{64}Co_{21}B_{15}$ amorphous
ribbon and an FeSi 7.8 wt.\% polycrystalline alloy (inset). The
theoretical prediction of Ref.~\cite{KUN-00} is also shown, with
$1/\sigma \nu z$ equal to $1.77$ and $2$, respectively, as given
by interface model \cite{DUR-00}.} \label{fig:sp}
\end{figure}

\begin{figure}
\centerline{
        \epsfxsize=7.5cm
                \epsfbox{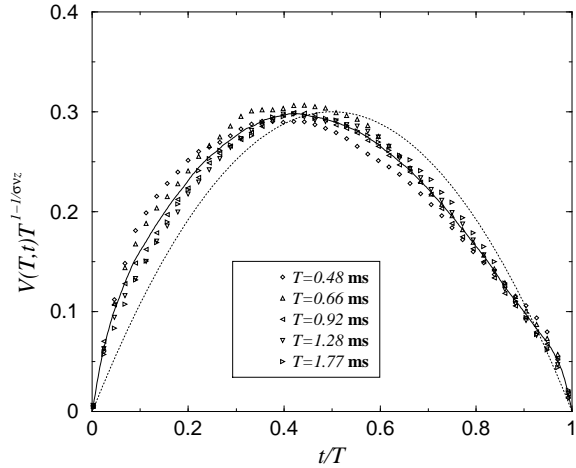}
        }
\caption{Average avalanche shape (eq.~\ref{Vt}) for the
$Fe_{64}Co_{21}B_{15}$ amorphous ribbon using the scaling exponent
$1/\sigma \nu z$ equal to $1.77$. The full line is the average
of the curves and the dotted line is a symmetric parabola.}
\label{fig:VT}
\end{figure}

\begin{figure}[htb]
\centerline{
        \epsfxsize=7.5cm
                \epsfbox{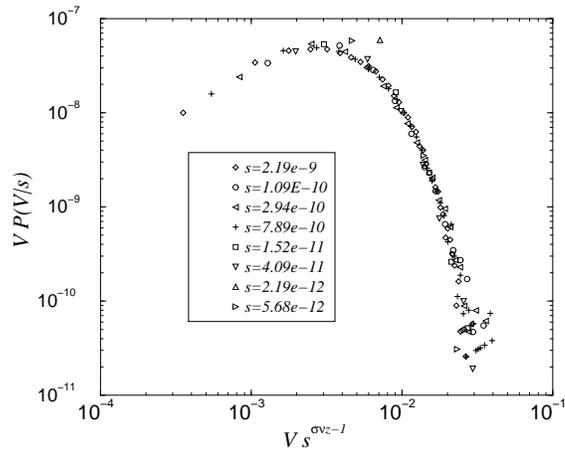}
        }
\caption{Distribution of avalanche voltages at fixed voltage
(eq.~\ref{PVS}) for the $Fe_{64}Co_{21}B_{15}$ amorphous ribbon
using the scaling exponent $1/\sigma \nu z$ equal to $1.77$. }
\label{fig:PVS}
\end{figure}

We consider two kind of samples, belonging to different
universality classes as pointed out in Ref.~\cite{DUR-00}. An
as-cast $Fe_{64}Co_{21}B_{15}$ amorphous alloy, measured under
moderate tensile stress ($\sigma \sim 20 MPa$), and an Fe-Si 7.8
wt.\% strip (30 cm $\times$ 0.5 cm $\times$ 60 $\mu$m) produced by
plan flow casting, having grains of average dimension of 25
$\mu$m. The amorphous ribbon follows the universality class where
the surface tension of the wall dominates the domain dynamics
(short range class), where $1/\sigma \nu z \sim 1.77$
\cite{DUR-00}. Fig.~\ref{fig:sp} shows the comparison of the power
spectrum with the average size distribution $\langle s(1/T)
\rangle$ as a function of the inverse of avalanche durations. The
agreement with the theoretical prediction is fairly good over an
extended time range: at high avalanche durations (small
frequencies), the time correlation between avalanches becomes
relevant and the theoretical analysis is no longer valid. The
inset of Fig.~\ref{fig:sp} show the same comparison in the case of
the FeSi sample. This material falls in the universality class
where long range magnetostatic fields dominates the domain
dynamics, giving $1/\sigma \nu z \sim 2$ \cite{ZAP-98}. Also in
this case, the agreement is fairly good, but for a smaller high
frequency range. The precise reason for this fact is not clear,
even though a visual inspection of the time signals of both
materials can justify this result: in the amorphous alloy, the
avalanches are well separated in time (see~\cite{DUR-99}), while
in the FeSi alloy the separation is much less defined
(see~\cite{DUR-95a}). This is also confirmed by the fact that only
in the latter material the critical exponents of size and duration
distributions strongly depend on the applied field rate. This
means that avalanche correlations, and thus time-time
correlations, are significantly different, giving different
spectral contributions. The results for the amorphous alloy are
confirmed by the scaling of the average avalanche shape $V(T,t)$
(fig.~\ref{fig:VT}) and of the probability $P(V|s)$
(fig.~\ref{fig:PVS}). In both figures the theoretical value
$1/\sigma \nu z \sim 1.77$ is used. Interestingly, the universal
scaling function $f_{shape}$ of eq.~\ref{Vt} is not an inverted
parabola as for the RFIM, but it shows a marked temporal
asymmetry. This is compatible with the results of
Ref.~\cite{PET-98a} concerning high order spectra: the high
frequency signal components precede the low frequency ones, so
that an avalanche of a given size and duration starts with a fast
signal ramp and relaxes at longer times. Interestingly enough,
despite this time asymmetry, the predictions of Ref.~\cite{KUN-00}
are confirmed, suggesting that the scaling properties are more
important than the exact shapes of average avalanches. This
conclusion strongly contradicts with the basic assumption often
reported in the literature \cite{MAZ-62b,ARQ-68}, where a
distribution of 'elementary' avalanches with a {\em pre-defined}
shape (often exponential) is summed up to calculate the power
spectrum.

We must add that the avalanches of the FeSi alloy does not show
such a nice scaling. In particular, short and long avalanches have
markedly different shapes, as the former is approximately an
inverted parabola, while the latter show a flat central region.
Also the scaling of $P(V|s)$ is not perfectly compatible with the
theoretical exponent $1/\sigma \nu z = 2$. All these
features could explain the limited agreement between the power
spectrum and $\langle s(T)\rangle$,
and surely need a more extensive analysis,
taking into account avalanche correlations and the dominant role
of demagnetizing fields.

From the  analysis of experimental results shown above, one may
argue that materials belonging to the short range class also
exhibit power spectra scaling as $1/\sigma \nu z$. As a matter of
fact, the application of larger applied tensile stresses on the
amorphous material does not change the universality class
\cite{DUR-99}, but reveals a more complex behavior. In particular,
$S(1/\omega)$ and $\langle s(T) \rangle$
do not longer scale in a similar way at
high stresses, even if $V(T,t)$ and $P(V,s)$ still rescale
approximately with $1/\sigma \nu z = 1.77$. This behavior reflects
the change of the avalanche shape as shown in Ref.~\cite{DUR-99}:
the cutoff of avalanche duration decreases while keeping the size
distribution invariant. As the scaling range of duration
distribution gets shorter, we expect that the longest avalanches,
close to the cutoff value, are increasingly more effective in the
time-time correlation of the signal, and thus to the power
spectra. Their frequency content is not obviously taken into
account in the scaling calculations of \cite{KUN-00}.

As clearly shown above, many different aspects can enter in the
definition of the scaling properties of the Barkhausen signal. As
already pointed out \cite{PET-98a,PET-98}, a more complex analysis
is required to evaluate in detail the statistics of avalanches.
In this respect, the analysis of Ref.~\cite{KUN-00} not only gives a
new approach to the long standing problem of power spectra shape,
but introduces precise statistical tools helpful both to
the experimental analysis and to the theoretical description of magnetization
processes. Within some limitations to be further investigated, the
results described above are fully compatible with our interface
models \cite{CIZ-97,ZAP-98,DUR-00}.
This is quite surprising, considering that this model
considers the motion of a single domain wall with strong
simplifications of the fields relevant to the dynamics (especially
of magnetostatic fields). In fact, a real material typically
shows a complex pattern of multiple domains with no easily
predictable field configuration. Here temporal and spatial
correlations are important, resulting in a very complex behavior
of time-time correlation and thus of power spectra.

\clearpage

\vspace{5mm}

\end{document}